\documentclass[nohyper,11pt,letterpaper]{JHEP3}

\usepackage{amsmath}
\usepackage{amsfonts}
\usepackage{amssymb}

\usepackage{graphics}
\usepackage{amsmath}
\usepackage{amsfonts}
\usepackage{amssymb}
\usepackage{graphicx}
\usepackage{epsfig}

\def\ba{\begin{eqnarray}}
\def\ea{\end{eqnarray}}
\def\be{\begin{equation}}
\def\ee{\end{equation}}

\def\d{\mathrm{d}}

\def\({\left(}
\def\){\right)}

\Roman{section}

\title{Scale-invariance in expanding and contracting universes
from two-field models }

\author{Andrew J. Tolley$^1$, Daniel H. Wesley$^2$ \\
~$^1$ Perimeter Institute for Theoretical Physics, Waterloo ON, N2L 2Y5, Canada.\\
~~~E-mail: \email{\tt atolley@perimeterinstitute.ca}\\
~$^2$ Department of Applied Mathematics and Theoretical Physics,
Cambridge University,\\
~~~Wilberforce Road, Cambridge CB3 OWA, United Kingdom.\\
~~~E-mail: \email{\tt D.H.Wesley@damtp.cam.ac.uk}\\
}

\abstract{We study cosmological perturbations produced by the most general two-derivative actions involving two scalar fields,
coupled to Einstein gravity, with an arbitrary field space metric, that admit scaling solutions. For contracting universes, we
show that scale-invariant adiabatic perturbations can be produced continuously as modes leave the horizon for any equation of
state parameter $w \ge 0$. The corresponding background solutions are unstable, which we argue is a universal feature of
contracting models that yield scale-invariant spectra. For expanding universes, we find that nearly scale-invariant adiabatic
perturbation spectra can only be produced for $w \approx -1$, and that the corresponding scaling solutions are attractors. The
presence of a nontrivial metric on field space is a crucial ingredient in our results.}

\begin{document}


\section{Introduction}

Single-field, slow-roll inflation has become the dominant paradigm for generating primordial density perturbations in the early universe. Despite its successes, it is still important to search for alternative models, if only to
determine whether the predictions of inflation are truly unique. Any proposed alternative must explain the nearly scale-invariant adiabatic primordial density fluctuations that seed large scale structure formation, and whose imprint is seen directly in the cosmic microwave background.

Models with a contracting phase, where density perturbations are generated before the traditional big bang, provide an interesting class of alternatives to inflation.
Proposals of this type include the recent ekyprotic and cyclic models
\cite{Khoury:2001wf,Steinhardt:2002ih}, and the older pre-big bang scenario \cite{Gasperini:2002bn}.  The ekpyrotic and cyclic
models create primordial cosmological perturbations during  a $w>1$ contracting phase, which has some similarity to an expanding
inflationary phase \cite{Erickson:2003zm,Gratton:2003pe,Khoury:2003vb}. However, it is well known that a single scalar field in a
contracting $w>1$ phase cannot produce nearly scale-invariant curvature perturbations, for $\zeta$ is not sensitive to the nearly
scale-invariant growing mode of the Newtonian potential, but only to the decaying mode
\cite{Lyth:2001pf,Khoury:2001zk,Hwang:2001ga,Brandenberger:2001bs,Creminelli:2004jg}.
Several
approaches to dealing with this problem
have been proposed.
 One can analyze the perturbations in higher dimensions
\cite{Tolley:2003nx,McFadden:2005mq,Battefeld:2004mn} in the hope that higher-dimensional effects will mix the growing and
decaying modes. Alternatively one may hope that during the evolution through the bounce from contraction to expansion a similar
effect occurs \cite{Peter:2002cn,Martin:2001ue,Tsujikawa:2002qc,Bozza:2005wn,Bozza:2005xs,Bozza:2005qg}. Or, one can posit that a
spectrum of scale-invariant isocurvature modes are subsequently converted into adiabatic modes
\cite{Notari:2002yc,Lehners:2007ac,Buchbinder:2007ad,Creminelli:2007aq,Koyama:2007mg}, an idea which has also been applied in
expanding universes \cite{Lyth:2002my}. In still other models, fluctuations are adiabatic and scale-invariant at horizon crossing
\cite{DiMarco:2002eb}.

In this work we describe a mechanism that generates scale-invariant spectra in both expanding and contracting universes.  We
consider the entire family of scaling solutions with two scalar fields, including a nontrivial metric on field space.  These
solutions serve as models of realistic systems in which the equation of state parameter $w$ is slowly varying, and are useful
tools for understanding the predictions of inflationary- and ekpyrotic-type models. We find that using our new two-field
solutions it is possible to construct contracting universe models in which a scale-invariant spectrum of adiabatic fluctuations is continuously produced at horizon crossing for any equation of state
parameter $w \ge 0$. The price is an instability in the background solution, which we show is universally present in all
two-field contracting universe models that produce scale-invariant spectra.
There are also inflationary solutions with nearly scale-invariant adiabatic perturbation spectra, but only when $w \approx -1$.
The inflationary background solutions are all stable to small perturbations.
Within the class of two scalar field models that admit scaling solutions, both the stability of the background solution and the
spectral indices of long-wavelength perturbations are determined by only two parameters. The first is the equation of state
parameter $w$ for the background solution, which we here parameterize by $c$, with $c=\sqrt{3(1+w)}$. The second parameter, which
we call $\Delta$, is constructed from the first and second derivatives of various functions that appear in the general action for
scaling solutions.

We will justify our claims with detailed calculations presented in the following sections.  Nonetheless, the essence of our
results can be understood using simple arguments.  The first important fact is that the behavior of fluctuations on sub-horizon
scales, with comoving wavenumbers $k > aH$, is determined by the behavior of $a(\eta)$ alone. This is because the action for the
curvature perturbation $\zeta$ is of the form (here $\eta$ is conformal time and $'$ denotes conformal time derivatives) \be S =
\int \frac{c^2 a(\eta)^2}{2} \left( (\zeta'_k)^2 - k^2\zeta^2_k + \cdots \right) \; \mbox{d}\eta,\label{eq:ZetaAction} \ee where
we have specialized to a single Fourier mode $\zeta_k(\eta)$ of comoving wavenumber $k$. Consequently, the standard boundary
conditions on $\zeta_k(\eta)$ are such that when well within the horizon each (positive-frequency) mode is
\be\label{eq:HeuristicSubHorizon} \zeta_k^+(\eta) = \frac{e^{-ik\eta}}{c\sqrt{2k}\, a(\eta)} \quad \text{for} \quad |k\eta| \gg
1. \ee  For scaling solutions, the equation of state parameter $w$ is constant, so $a(\eta) \propto |\eta|^{2/(1+3w)}$. At
horizon crossing, where $|k\eta| = 1$, the amplitude of each  mode is thus \be\label{eq:HeuristicHorizonCrossing} \zeta_k(\eta) =
\tilde{C} k^B \quad \text{with} \quad B=\frac{3(1-w)}{2(1+3w)} \quad \text{when} \quad |k\eta| = 1 ,\ee with $\tilde{C}$ a
$k$-independent constant.  The modes will only be exactly scale-invariant as they cross the horizon, with $B = -3/2$, when $w=
-1$.

In the absence of entropy perturbations, such as in single-inflaton models, it is well known that $\zeta_k' = 0$ outside the
horizon.  In this case, the only way to generate a scale-invariant curvature spectrum as modes leave the horizon in an expanding
phase is with $w \simeq -1$, \emph{i.e.} inflation. In a contracting universe with $w>1$,  $\zeta$ is conserved in single-field
models, so matching across the horizon cannot give a scale-invariant spectrum. When $-1/3<w<1$, one finds that $\zeta$ grows
outside the horizon as $\zeta_k\approx |\eta|^{3(w-1)/(1+3w)}$, which can be used to give $\zeta_k$ a scale-invariant spectrum if
$w=0$ \cite{Finelli:2001sr,Wands:1998yp}. However, the $w=0$ background possesses an instability, and this mechanism is also
problematic since a $w=0$ component in a contracting universe will be quickly overtaken by radiation and anisotropy as the scale
factor shrinks to zero \cite{Erickson:2003zm}.

When entropy perturbations are present, such as in multi-inflaton models, then $\zeta_k'$ need not vanish. It is then possible to
continuously generate scale-invariant adiabatic perturbations at horizon crossing if $\zeta_k$ evolves in the correct way on
super-horizon scales. Specifically, suppose $\zeta_k$ evolves as \be \zeta_k(\eta) \sim |\eta|^p \quad \text{for} \quad |k\eta|
\ll 1, \ee for some exponent $p$.  Then using (\ref{eq:HeuristicHorizonCrossing}) one finds that $\zeta_k$ will develop an
exactly scale-invariant spectrum provided that \be\label{eq:RequiredZetaGrowthExponent} p = - \frac{3(1+w)}{1+3w}. \ee If we are
free to adjust $p$ at will, the only remaining requirement is that modes are exiting the horizon during the time when primordial
perturbations are generated.  This requires that $w > -1/3$ for contracting solutions, and $w < -1/3$ for expanding solutions. We
will see below that once $w$ is fixed in the two-field scaling solutions, varying $\Delta$ provides the freedom to adjust $p$ as
required for the contracting solutions, provided that in addition that $w\ge 0$. A special case of this mechanism was considered
in \cite{DiMarco:2002eb}. In the expanding case, the growing mode always has $\zeta$ constant and this mechanism cannot be
applied.

The full analysis shows that even for the contacting case this mechanism cannot be implemented for scaling solutions when the
metric on field space is Euclidean.  That is, if there exists a field redefinition that takes the action for the scalar fields
$\Phi^a$ into the form \be S = -\int \frac{a(\eta)^2}{2} \left( [\partial\Phi_1]^2 + [\partial\Phi_2]^2 + \cdots \right) \;
\mbox{d}^4 x, \ee then we will show that $\zeta$ is constant on super-horizon scales,   despite the presence of entropy
perturbations.  The underlying reason is that adiabatic and entropy perturbations decouple for for scaling solutions with
Euclidean field space metrics, and so the usual arguments regarding single-field adiabatic perturbations will apply
\cite{Lyth:2001pf,Creminelli:2004jg}. Therefore, in these cases the only possibility for generating scale-invariant adiabatic
perturbations in $\zeta_k$ is an expanding nearly de Sitter solution with $w\simeq -1$ or a contracting solution with $w \simeq
0$.  The mechanisms proposed recently by \cite{Lehners:2007ac}, and also employed in \cite{Buchbinder:2007ad,Creminelli:2007aq},
rely on mixing of isocurvature into adiabatic modes by turning a corner in field space, during which time the solution is far
from scaling for a brief period. Here, the non-trivial metric on field space provides the crucial coupling between adiabatic and
entropy perturbations on super-horizon scales.

Ideally, the background solutions used to obtain these perturbation spectra would be stable, and so fine-tuned
initial conditions would not be required.  For contracting universes, the heuristic arguments given above hint that this will not be possible.  For these solutions, which have $w
> 1$, the expression (\ref{eq:RequiredZetaGrowthExponent}) implies that we require $p<0$, and so $\zeta_k$ grows on super-horizon
scales.  The growth of long-wavelength modes suggests an instability in the homogeneous degrees of freedom of the system.  We
will explicitly verify this instability in the scaling solutions by studying the fixed points of the autonomous dynamical system
associated with the homogeneous field and metric modes.  We will also argue that this instability must always be present if a
scale-invariant spectrum is to be created in a contracting universe -- even if it is not initially created in the adiabatic
perturbation modes.  Whether this instability is really a problem depends on the specific model under consideration.  The
instability would only be present during the time in which the background approximately corresponds to a scaling solution, and
would in any event be naturally cut off by the bounce itself.  In the expanding case, where $-1 \le w < -1/3$, the expression
(\ref{eq:RequiredZetaGrowthExponent}) implies that $p \ge 0$ and so $\zeta_k$ is either constant or decays on super-horizon
scales in order to produce a nearly scale-invariant spectrum.  This suggests that the inflationary background solutions are
stable, which we verify by studying the associated autonomous system.

Inflation models with multiple fields have been a fertile research topic for some time: see for example
\cite{Kofman:1985aw}-\cite{vanTent:2003mn}. Some basic ingredients of the mechanism we describe here, such as the generation of
isocurvature perturbations or non-flat metrics on field space, have been studied before in the context of slow-roll inflation.
Our novel contribution is to discard the slow-roll assumption by concentrating on scaling solutions, which allows for an exact
calculation of the spectral indices in terms of $(c,\Delta)$. Our results indicate that it is still necessary to be close to
slow-roll, i.e. have $w \approx -1$ to achieve an approximately scale-invariant spectrum. In collapsing models there is no
natural slow-roll condition, and models with multiple fields that produce scale-invariant isocurvature fluctuations at horizon
crossing, which are later converted to adiabatic fluctuations, have attracted a great deal of attention recently
\cite{Lehners:2007ac,Buchbinder:2007ad,Creminelli:2007aq,Koyama:2007mg}.  In the present work we directly produce adiabatic
fluctuations at horizon crossing, thanks to the non-Euclidean metric on field space.

The scaling solutions that we employ in this analysis are the most general for two-derivative actions with two fields.  We
include an arbitrary metric on field space, but otherwise the kinetic terms are canonical.  Once the equation of state parameter
$w$ is fixed for the background solution, the relevant action is completely specified by three additional free parameters
$f_1,f_2$ and $h_2$, which arise from the expansion to quadratic order of some free functions appearing in the scaling action.  A
remarkable feature of these scaling solutions is that their stability properties and the spectral index of the perturbations
depend entirely on $c = \sqrt{3(1+w)}$ and on a single additional parameter, which we call $\Delta$, that is itself constructed
from $f_1,f_2$ and $h_2$.  This simple dependence makes it quite straightforward to study the phenomenology of the two field
model, as well as to include a variety of other models studied by previous researchers within the framework described here. In
these senses we have characterized the natural two-field two-derivative generalizations of the single-field scaling solutions which have been studied in
both expanding and collapsing universes \cite{Lyth:1991bc,Gratton:2003pe,Boyle:2004gv}.

In the present work we do not attempt to embed our mechanism within a complete model.  Instead, we focus on questions of
principle, and endeavor to show that it is possible to construct models with a simple field content and Lagrangian that will
produce certain types of primordial perturbation spectra.  We find that the combination of including a curved metric on field is
enough to drastically modify the predictions from the single-field case for contracting models, but not for expanding ones.
Questions such as whether the initial conditions for the contracting solutions are truly fine-tuned will depend on the details of
specific models. The relative amplitudes of the curvature and isocurvature perturbations after the universe enters radiation
domination will also depend on the details of reheating after inflation (or its analogue in contracting universe models).  We
will not consider the details of the process by which inflation ends in expanding models, or the big crunch/big bang transition
in contracting models.

As in other contracting universe models, the mechanism described here produces its own distinctive observational signature in the tensor spectrum. 
While the spectral index for scalar perturbations
depends on the detailed coupling between adiabatic and isocurvature modes on super-horizon scales, the tensor spectrum does not.
The tensor modes decouple from everything but the background metric, and so they directly probe the expansion history (and
therefore $w$) during the time in which the perturbations are generated. The kinetic part of the action for the tensor
perturbation $h_k$ is given by (\ref{eq:ZetaAction}) under the replacement $c\zeta_k \leftrightarrow h_k$.  The tensor mode $h_k$
evolves like (\ref{eq:HeuristicSubHorizon}) while sub-horizon and is constant when super-horizon, so by
(\ref{eq:HeuristicHorizonCrossing}) the tensor spectral index $n_T$ is \be n_T = 3 + 2B = 6 \left(\frac{1+w}{1+3w}\right). \ee
{
In contracting models, the
tensor spectrum remains quite blue, with $2 > n_T > 3$. The more general solutions we study are therefore very relevant for
searches for primordial gravitational waves in the cosmic microwave background or with laser interferometers.

This paper is organized as follows.  In Section \ref{s:ScalingSolutions} we describe the class of two scalar field models that
admit scaling solutions.  Any such model can be completely described by four parameters $c,f_1,f_2$, and $h_2$.  We also give
explicit examples, one with a Euclidean field space metric, and one in which field space is curved.  In Section
\ref{s:Perturbations} we set up and solve the cosmological perturbation equations for the scaling backgrounds.  We show that the
spectral indices of isocurvature and adiabatic modes depend only on $c$ and one additional parameter $\Delta$, which is
constructed from  $c,f_1,f_2$, and $h_2$.  We also show how super-horizon isocurvature and adiabatic modes decouple when the
field space metric is Euclidean.  Section \ref{s:Stability} is devoted to an analysis of the stability of the background scaling
solutions, which is also completely controlled by $c$ and $\Delta$.   We conclude in Section \ref{s:Conclusions}.

\section{Scaling solutions}\label{s:ScalingSolutions}

A scaling solution is a solution to the equations of motion for which all contributions to the total energy
density scale identically with time, keeping their fractional contributions constant. Although the system is evolving in time, there is a
sense in which it is time translation invariant. To be precise, scaling solutions arise when there exists a timelike homothetic Killing vector\footnote{A homothetic Killing vector $\xi^a$ satisfies $\nabla_a\xi_b+\nabla_b \xi_a=C g_{ab}$ where $C$ is a constant.} on spacetime, \emph{i.e.} a scale invariance.

Consider the action for a set of scalar fields $\Phi^a$ with a nontrivial field space metric $G_{ab}(\Phi)$ coupled to gravity\footnote{We shall work in units where $M_{pl}^{-2}=8\pi G=1$.}
\be S=\int \(\frac{1}{2}R-\frac{1}{2}G_{ab} \, g^{\mu\nu}\partial_{\mu}\Phi^a
\partial_{\nu}\Phi^b -V(\Phi)\) \, \sqrt{-g}\; \d^4 x.
\ee A scaling solution exists when there is a continuous transformation with parameter $\lambda$ such that \be\label{eq:ScalingConditions}
\frac{\d\Phi^a}{\d\lambda} = \xi^a(\Phi) \quad g_{\mu\nu} \rightarrow e^\lambda \, g_{\mu\nu} \quad S \rightarrow e^\lambda \, S.
\ee
 Any such transformation preserves the form of the
equations of motion\footnote{A slightly different approach to multi-field scaling solutions has been considered in
\cite{Karthauser:2006ix}.}. These conditions imply that $\xi^a$ is a Killing vector on field space, and that the potential
transforms as \be V \to e^{-\lambda}V. \ee If we choose coordinates on field space so that the Killing direction is $\phi$, and
$\sigma^1, \dots, \sigma^n$ are the other directions, then we have sufficient coordinate freedom to put the metric in the form
\be G_{ab}\partial \Phi^a
\partial \Phi^b = f(\sigma^1, \dots, \sigma^n) (\partial \phi)^2+2 A_{i}(\sigma^1, \dots, \sigma^n)\partial \sigma^i \partial
\phi+h_{ij}(\sigma^1, \dots, \sigma^n)\partial \sigma^i
\partial \sigma^j, \ee and the potential in the form \be V=V_0 \exp(-c \phi) h(\sigma^1, \dots, \sigma^n). \ee Here $A_{i}$ is
only defined up to $A_{i}\rightarrow A_{i}+\frac{\partial \chi}{\partial \sigma^i}$, corresponding to redefinitions $\phi
\rightarrow \phi+\chi(\sigma^1 \dots \sigma^n)$, with an associated redefinition of $h_{ij}$. In the two field case ($n=1$) we can use
this freedom to set $A_1=0$.
The metric then satisfies the conditions (\ref{eq:ScalingConditions}) with a Killing vector given by $\xi^a=(1/c,0, \dots,0)$.
In
these coordinates, the scaling solution has only $\phi$ changing with time, and
$\sigma^1, \dots \sigma^n$ are constants which can be taken to be zero by an appropriate coordinate transformation. It is then
convenient to rescale $\phi$ so that $f(0, \dots,0)=1$ and rescale $V_0$ so that $h(0,\dots,0)=1$. The net result is that the
scaling solution is just the usual single field scaling solution associated with the action \be S_{equiv}=\int
\(\frac{1}{2}R-\frac{1}{2}(\partial \phi)^2 -V_0 \exp(-c \phi)\) \,\sqrt{-g} \, \d^4 x . \ee Nevertheless perturbations around
this solution will be very different, since fluctuations in $\phi$ and the $\sigma^i$ will probe the terms in the Taylor
expansions of $f(\sigma^i)$ and $h(\sigma^i)$ about the background solution.

\subsection{The general case}

The action for any two-field system that admits scaling solutions can be put in the form \be S=\int
\(\frac{1}{2}R-\frac{1}{2}f(\sigma) (\partial \phi)^2-\frac{1}{2} (\partial \sigma)^2 -V_0 \exp(-c \phi)h(\sigma)\)\;  \sqrt{-g} \, \d^4x. \label{originalaction} \ee There is
sufficient coordinate freedom to set $\sigma=0$ in the background solution, so for linearized perturbations we will only need
the second order Taylor expansions of the two
functions \ba f(\sigma)&=&1+f_1 \sigma+ f_2 \sigma^2+\dots, \\
h(\sigma)&=&1+h_1 \sigma+ h_2 \sigma^2+\dots. \ea This is a useful simplification since the result will depend on only a handful
of coefficients.
In fact, for $\sigma=0$ to be a consistent solution we must have
\be
h_1=-c^2 f_1/(c^2-6).
\ee
This is not a physical
restriction on the choice of potential and kinetic term, but instead reflects our choice of coordinates on field space.
The coefficient $V_0$ is degenerate with shifts $\phi \to \phi + C$, but we will leave this symmetry unfixed.
The net result is that
the free parameters in the Lagrangian are $(c,f_1,f_2,h_2)$.

\subsection{A Euclidean example}

To illustrate the above analysis it is useful to look at an example recently considered by Lehners {\it{et al.}}
\cite{Lehners:2007ac} in the context of the ekpyrotic model (see also \cite{Liddle:1998jc,Malik:1998gy,Finelli:2002we} in the
context of assisted inflation), of two minimally coupled fields with exponential potentials \be S=\int \(\frac{1}{2}R-\frac{1}{2}
(\partial \Phi_1)^2-\frac{1}{2} (\partial \Phi_2)^2 -V_0 \exp(-c_1 \Phi_1)-V_0 \exp(-c_2 \Phi_2)\) \,\sqrt{-g}\,\d^4 x  , \ee
where we have used one of the shift symmetries $\Phi_j \to \Phi_j + c_j$ to equalize the coeffients in front of the exponentials.
In these coordinates the scaling solution has \be \Phi_j(\eta) = A_j \ln (\eta) + B_j, \ee with $c_1 A_1 = c_2 A_2$.
We can perform a rotation on field space to define \ba
\phi   &=& \frac{c_2 \Phi_1 + c_1 \Phi_2}{\sqrt{c_1^2 + c_2^2}}, \\
\sigma &=& \frac{c_1 \Phi_1 - c_2 \Phi_2}{\sqrt{c_1^2 + c_2^2}} + \sigma_0, \ea so that the action becomes
\be
S=\int
\(\frac{1}{2}R-\frac{1}{2} (\partial \phi)^2-\frac{1}{2} (\partial \sigma)^2 -\tilde{V}_0 \exp(-c \phi) h(\sigma)\)
\,\sqrt{-g}\,\d^4 x,
\ee
where
\be
\frac{1}{c^2}=\frac{1}{c^2_1}+\frac{1}{c^2_2},
\ee
and we used the second shift symmetry to
set
\be
\sigma_0 = \frac{2 \ln (c_2/c_1)}{c_1^2 + c_2^2}
\ee
 which gives
\be
\tilde V_0 = \left[ \(\frac{c_2}{c_1}\)^\frac{2c_1^2}{c_1^2+c_2^2} +
\(\frac{c_1}{c_2}\)^\frac{2c_2^2}{c_1^2+c_2^2}\right] V_0
\ee
  and
\be
h(\sigma) = 1 + \frac{c^2}{2}\sigma^2 + \cdots
\ee
so that $h_2 = c^2/2$ but $f_1=f_2=h_1=0$.
The scaling solution considered in \cite{Lehners:2007ac} corresponds to
\be
\phi = \phi_0 \ln(\eta)
\ee
and $\sigma=0$.
This case is very special since the field space metric is
Euclidean and so there are no metric couplings.

\subsection{A curved example}

As an example of a two-field model with a nontrivial field space metric which still exhibits scaling solutions, consider the
following action \be S=\int \(\frac{1}{2}R-\frac{1}{2}\(\frac{(\partial b)^2+(\partial \psi)^2}{\lambda^2 b^2}\)-V_0
\frac{b^{\alpha}\psi^{\beta}}{b_0^{\alpha}\psi_0^{\beta}}\) \, \sqrt{-g}\, \d^4x . \ee The field space metric is a two
dimensional hyperboloid, with constant negative curvature. This kinetic term arises often in string compactifications, where it
is more common to define $z=\psi+ib$ so that the kinetic term is \be \frac{(\partial b)^2+(\partial \psi)^2}{\lambda^2 b^2} \to
\frac{\partial z\partial \bar{z}}{[\lambda {\rm Im}(z)]^2}. \ee Through the redefinition $b = \lambda^{-1} e^{-\lambda \chi}$ the
kinetic term is cast into another useful form \be \frac{(\partial b)^2+(\partial \psi)^2}{\lambda^2 b^2} \to (\partial \chi)^2 +
e^{2\lambda\chi}(\partial \psi)^2, \ee which has been studied a number of times in the past. Returning to the original variables
$(b,\psi)$, we seek scaling solutions with equation of state parameter $w$, having \be \frac{b(\eta)}{b_0}
=\frac{\psi(\eta)}{\psi_0} = \(\frac{\eta}{\eta_0}\right)^{3A(w-1)/(1+3w)} \ee and $a(\eta) \propto |\eta|^{2/(1+3w)}$.  These
solutions exist provided that \ba
A^2 &=& \frac{4\lambda^2}{3}\frac{1+w}{(1-w)^2}\(1+\left[\frac{\psi_0}{b_0}\right]^2\)^{-2}, \\
\alpha &=& \frac{2}{A}\frac{1+w}{1-w}\frac{1-[\psi_0/b_0]^2A}{1+[\psi_0/b_0]^2},\\
\beta &=& \frac{2}{A}\frac{1+w}{1-w}\left[\frac{\psi_0}{b_0}\right]^2
\frac{1+A}{1+[\psi_0/b_0]^2},
\ea
and so if $(\alpha,\beta)$ are freely adjustable then the
scaling solutions are specified by $(w,\lambda,\psi_0/b_0)$.
The parameter $\psi_0/b_0$ is essentially the ratio of the kinetic energies
of the two fields. With the field redefinition
\ba
b&=& \frac{2Ce^{\lambda \sigma-\kappa \phi}}{\psi_0(1+C^2 e^{2\lambda \sigma})},\\
\psi&=& \frac{e^{-\kappa \phi}(C^2 e^{2\lambda \sigma}-1)}{\psi_0(1+C^2e^{2\lambda \sigma})},\ea
 the action becomes
\be S=\int \d^4 x \sqrt{-g} \(\frac{1}{2}R-\frac{1}{2}f(\sigma) (\partial \phi)^2-\frac{1}{2} (\partial \sigma)^2 -V_0 e^{-c\phi}
h(\sigma)\), \ee where \ba f(\sigma)&=&
\frac{(e^{-\lambda \sigma}+C^2 e^{\lambda\sigma})^2}{(1+C^2)^2},\\
\\
h(\sigma) &=&
\left[ \frac{(1+C^2)e^{\lambda\sigma}}{1+C^2e^{2\lambda\sigma}}\right]^\alpha \left[ \frac{C^2+1}{C^2e^{2\lambda\sigma}+1}
\frac{C^2e^{2\lambda\sigma}-1}{C^2-1}\right]^\beta \ea To obtain these expressions we have applied the requirement that the
scaling solution corresponds to $\sigma=0$ and that $f(0)=h(0)=1$, which fixes \be \kappa^2 = \frac{4C^2 \lambda^2}{(1+C^2)^2}.
\ee As we will see below, the quantity that determines whether isocurvature and adiabatic perturbations are coupled on
super-horizon scales is $f_1$, which in this case is given by \be f_1 = 2\lambda\left(\frac{C^2-1}{C^2+1}\right). \ee The
variable $C$ parameterizes the family of scaling solutions with fixed $w,\lambda$.  In terms of the ratio of kinetic energies, we
have the relation \be \frac{\psi_0}{b_0} = \frac{1}{2} \left( C - \frac{1}{C} \right). \ee For the special value $C=1$, the
coupling parameter $f_1$ vanishes.  This special value of $C$ corresponds to $\psi_0/b_0 = 0$, which is the case in which  there
is no kinetic energy in the $\psi$ field.  The quantities $f_2,h_1$ and $h_2$ can be readily calculated from the expressions for
$f(\sigma)$ and $h(\sigma)$.

\section{Two-field perturbations}\label{s:Perturbations}

In this section we analyze the perturbations produced by the two field system in uniform-field gauge $\delta \phi=0$.  To connect
quantities in this gauge to gauge-invariant variables, we define the conventional gauge-invariant adiabatic perturbation variable
$\zeta$ (also called $\mathcal {R}$) as
\be
\zeta=\psi+
\frac{a'}{a}\frac{\delta\phi}{\phi'},
\ee where $\psi$ is the spatial metric perturbation, and so in uniform-field gauge $\zeta=\psi$. The fluctuation
$\delta \sigma$ is automatically gauge invariant since
 $\sigma=0$ in the background.
To stress the connection to gauge-invariant quantities we express the perturbations directly in terms of $\zeta$ and $\delta
\sigma$. Had we chosen to use coordinates on field space where both fields are varying (e.g. the $\Phi_1$ and $\Phi_2$ considered
above), a more complex projection would be required to separate the adiabatic and isocurvature modes (see for example
\cite{Gordon:2000hv} for an exposition of this technique). Our analysis is valid for both contracting (ekpyrotic) and expanding
(inflationary) solutions through an appropriate choice of the equation of state $w$.

\subsection{Background scaling solution}

The background equations of motion are
 \ba
3\left(\frac{a'}{a}\right)^2 &=& \frac{f}{2} \phi'^2 + \frac{1}{2}\sigma'^2 + a^2 V_0 e^{-c\phi} h , \label{eq:BackgroundEOM1}\\
0&=& \phi'' + 2\frac{a'}{a} \phi' + f^{-1} \frac{df}{d\sigma} \phi' \sigma' - ca^2V_0 e^{-c\phi} f^{-1}h,  \\
0&=& \sigma'' +   2\frac{a'}{a} \sigma' - \frac{1}{2} \frac{df}{d\sigma} \phi'^2 + a^2 V_0 e^{-c\phi} \frac{dh}{d\sigma}.
 \label{eq:Friedmann1}\label{eq:BackgroundEOM3}
\ea The background scaling solution expressed in terms of conformal time is
\ba
a(\eta) &=& (\eta/\eta_0)^{\hat{p}},\\
\phi &=&\phi_0+\frac{2c}{c^2-2}\ln (\eta/\eta_0), \\
\sigma &=&0, \ea where $\hat{p}=2/(1+3w)$ and $w=c^2/3-1$ is the equation of state parameter. The parameter $\phi_0$ is given by
\be \phi_0=\frac{1}{c}\ln\(\frac{-(c^2-2)^2\eta_0^2 V_0}{2(c^2-6)}\). \ee Note that $V_0>0$ for $c^2<6$ and $V_0<0$ for $c^2>6$.
Consistency of the solution $\sigma=0$ with (\ref{eq:Friedmann1}) enforces $h_1=-c^2 f_1/(c^2-6)$.

\subsection{Uniform-field gauge perturbations}

The calculation of the two field perturbations is straightforward to perform in uniform-field gauge. The scalar metric
perturbations are defined via \be \d s^2=a(\eta)^2\(-(1+2\Phi)\d\eta^2+2 \nabla_i \alpha \d x^i \d\eta+(1-2 \zeta)\d\vec{x}^2 \),
\ee where we have used the residual gauge freedom to set anisotropic parts of the 3-metric to zero. $\Phi$ and $\alpha$ are
essentially the lapse and shift perturbations, respectively, in the ADM decomposition. They can be uniquely determined in momentum space in
terms of the physical degrees of freedom by solving the two constraint equations \ba
\delta G^0_0=\delta T^0_0 &\rightarrow& \alpha_k= \frac{2c^2f_1\delta \sigma_k-2(c^2-6)\Phi_k+(c^2-2)((c^2-2)k^2\eta^2\zeta_k+6\eta \zeta'_k)}{2(c^2-2)k^2\eta},\\
\delta G^0_i=\delta T^0_i &\rightarrow& \Phi_k=-\frac{1}{2}(c^2-2)\zeta_k' .\ea The remaining Einstein equations are redundant as
a consequence of the Bianchi identities, and so the equations of motion for the two physical degrees of freedom can be obtained
directly from the two perturbed equations for the scalar fields \ba &\zeta_k''&+\frac{4}{(c^2-2)\eta}\zeta_k'+k^2 \zeta_k =
\frac{2(c^2-6)f_1}{(c^2-2)^2\eta^2} \delta \sigma_k -\frac{2f_1}{(c^2-2)\eta}\delta \sigma_k', \label{eq:pert1}
 \\
 &\delta \sigma_k''&+\frac{4}{(c^2-2)\eta}\delta \sigma_k'+k^2 \delta \sigma_k +\frac{(24 h_2-4c^2(f_2+h_2))}{(c^2-2)^2\eta^2}\delta \sigma_k = \frac{2c^2 f_1}{(c^2-2)\eta} \zeta'_k.
\label{eq:pert2} \ea These equations of motion arise from the quadratic action $S_{(2)}$, obtained by expanding the original action (\ref{originalaction}) to second order and re-expressing it in terms of our perturbation variables $\zeta_k$, $\delta \sigma_k$, which gives
\ba S_{(2)} = \int & \Bigg( &
\frac{1}{2}(\zeta_k')^2+\frac{1}{2c^2}(\delta \sigma_k')^2-\frac{1}{2}k^2\zeta_k^2-\frac{1}{2c^2}k^2\delta \sigma_k^2 \nonumber\\
&-& \frac{(24 h_2-4c^2(f_2+h_2))}{c^2(c^2-2)^2\eta^2}\delta \sigma_k^2+\frac{2f_1}{(c^2-2)\eta}\delta \sigma_k \zeta'_k \Bigg) \,
c^2 a(\eta)^2 \, \d \eta . \label{eq:actionpert} \ea The nature of the solutions depends crucially on the kinetic coupling
parameter $f_1$. If $f_1=0$ then the adiabatic and isocurvature modes decouple: this is the case when the field space metric is
Euclidean. If $f_1 \neq 0$ then the adiabatic and isocurvature modes are inextricably coupled. That is, we have already used all
of our reparameterizations to put the action in the form (\ref{eq:actionpert}), so the coupling term cannot be removed. Because
of the significant differences between the $f_1=0$ and $f_1 \ne 0$ cases, in what follows we shall consider the two solutions
separately.

The inner product of two sets of modes can be inferred from the action (\ref{eq:actionpert}), and is given in momentum space by
the analogue of the Klein-Gordon norm \ba \big\langle (\tilde{\zeta}_k , \delta \tilde{\sigma}_k),(\zeta_k,\delta \sigma_k) \big\rangle =i
a^2(\eta)
\bigg(  & c^2 & \tilde{\zeta}_k^* \zeta'_k -c^2 \tilde{\zeta'}_k^* \zeta_k+\delta \tilde{\sigma}_k^* \delta \sigma'_k \nonumber \\
&-& \delta \tilde{\sigma'}_k^* \delta \sigma_k+\frac{2f_1c^2}{(c^2-2)\eta} (\tilde{\zeta}_k^*\delta \sigma_k- \delta
\tilde{\sigma}_k^* \zeta_k) \bigg), \label{eq:innerproduct}\ea which is defined so that a positive frequency mode has unit norm.

\subsubsection*{Decoupled case: $f_1=0$}

In the decoupled case we can find exact solutions for the mode functions in terms of Hankel functions. The two independent
positive frequency modes for each field are given by
\ba {\bf I} &&\, \, \, \, \zeta^+_k = \frac{1}{c a(\eta)\sqrt{2k}}\sqrt{\frac{\pi}{2}}\sqrt{-k\eta}H_{\nu}^{(1)}(-k\eta) \, , \, \delta \sigma^+_k=0 \\
{\bf II} &&\, \, \, \, \delta \sigma^+_k =
\frac{1}{a(\eta)\sqrt{2k}}\sqrt{\frac{\pi}{2}}\sqrt{-k\eta}H_{\tilde{\nu}}^{(1)}(-k\eta) \, , \, \zeta^+_k=0 \ea where \be
\nu=\frac{c^2-6}{2(c^2-2)} ,\ee and \be \tilde{\nu}=\frac{\sqrt{(c^2-6)^2+16c^2(f_2+(1-6/c^2)h_2)}}{2(c^2-2)}.\ee
 Since the isocurvature and adiabatic modes are decoupled, they will have different long-wavelength spectral
indices. We define the power spectra in the usual way
\ba P_{\zeta}(k,\eta_f)&=&\frac{k^3}{2\pi^2}\langle\zeta_k^2(\eta_f)\rangle, \\
P_{\sigma}(k,\eta_f)&=&\frac{k^3}{2\pi^2}\langle\delta \sigma_k^2(\eta_f)\rangle ,\ea where $\eta_f$ is the time at which the
generation process ends, such as at reheating in the inflationary case. Since $\zeta$ is a constant outside the horizon,
$P_{\zeta}$ does not depend on $\eta_f$ and we find an adiabatic spectral index \be
n_s=4-2|\nu|=4-\frac{|c^2-6|}{|c^2-2|}.
\ee
 So as for a
single field the only two case which give scale invariance are $c^2=0$
\emph{i.e.} $w=-1$ expanding, or $c^2=3$, $w=0$ contracting. By contrast
$\sigma$ will in general be evolving outside the horizon and so its amplitude will depend on when the generation process ends.
The isocurvature spectral index of the modes which have already left the horizon are given by $n_s=4-2|\tilde{\nu}|$, which can
be freely adjusted by changing the coupling constants. This fact has already been used in many scenarios \cite{Lyth:2002my,Notari:2002yc,Lehners:2007ac,Buchbinder:2007ad,Creminelli:2007aq} to generate
scale invariant isocurvature modes which are subsequently converted to adiabatic ones by a second mechanism.

\subsubsection*{Coupled case: $f_1\neq 0$}

When their equations of motion are coupled, the isocurvature and adiabatic modes must have essentially the same dependence on
$\eta$.  This implies that both modes have the same spectral index at long-wavelengths. We compute the spectral index by taking
the usual WKB approximation within the horizon, matching this solution at horizon crossing with a power law super-horizon mode.
Because our solutions are scaling this will reproduce the \emph{exact} spectral index, but only give the amplitude approximately.

For $k \gg aH$ the positive frequency modes are given by their WKB form \ba \zeta_k^+=\frac{A}{a(\eta)\sqrt{2k}} \(-k\eta\)^s e^{-ik\eta}, \\
\delta \sigma_k^+ =\frac{B}{a(\eta)\sqrt{2k}} \(-k\eta\)^s e^{-ik\eta},
 \ea
 where
\be s_{\pm}=\frac{\pm if_1 c}{(c^2-2)}. \ee Consistency requires \be \frac{B}{A}=-\frac{(c^2-2)s}{f_1}=\mp i c. \ee Again we have
two independent pairs of solutions and we should distinguish between the two which are orthogonal with respect to
(\ref{eq:innerproduct}). In fact the above modes with the choice $s_+$ are orthogonal to those with $s_-$ and so the two
independent and properly normalized modes (positive frequency) are
\ba {\bf I} &&\, \, \, \, \zeta_k^+=\frac{1}{c a(\eta)\sqrt{4k}} \(-k\eta\)^{s_+} e^{-ik\eta} \, , \, \delta \sigma_k^+=\frac{-i}{a(\eta)\sqrt{4k}} \(-k\eta\)^{s_+} e^{-ik\eta} \\
{\bf II} &&\, \, \, \, \zeta_k^+=\frac{1}{c a(\eta)\sqrt{4k}} \(-k\eta\)^{s_-} e^{-ik\eta} \, , \, \delta
\sigma_k^+=\frac{+i}{a(\eta)\sqrt{4k}} \(-k\eta\)^{s_-} e^{-ik\eta} .\ea
The behavior of modes in the coupled and decoupled cases are essentially the same when well within the horizon:
compare the
above to (\ref{eq:HeuristicSubHorizon}). We cannot modify the contribution to the spectral index from the subhorizon physics.

For $k \ll aH$ the two modes must have the same power law dependence. Since in the $k\eta \to 0$ limit (\ref{eq:pert1}) and
(\ref{eq:pert2}) are a coupled system of Euler equations, we take  $\zeta_k= A' \eta^p$ and $\delta \sigma_k= B' \eta^p$ and
substitute into (\ref{eq:pert1}) and (\ref{eq:pert2}) giving\footnote{We would like to thank Krzysztof Turzynski for pointing out
a crucial oversight in an earlier version of
 this manuscript which modifies the previously stated conclusions.},
\be\label{eq:ABp_1} (6-c^2+p(c^2-2))(2B' f_1+A' (c^2-2)p)=0 ,\ee and \be\label{eq:ABp_2} 
B'(24h_2-4c^2(f_2+h_2))+(c^2-2)(-2A'c^2f_1+B' p(6-c^2+p(c^2-2)))=0. \ee
From the first equation we have one solution with
\be\label{eq:wsoln1} p=\frac{(c^2-6)}{c^2-2},
\ee 
for which the second equation implies
\be\label{eq:wsoln2} B'=
-\frac{A'c^2(c^2-6)f_1}{2(-6h_2+c^2(f_2+h_2))} .
\ee 
Returning to (\ref{eq:ABp_1}), the remaining solutions have
\be \label{eq:solnppmA}
B'=-\frac{A'(c^2-2)p}{2 f_1},
\ee and substituting into (\ref{eq:ABp_2}) gives
 \be
p(c^4(p-1)p+4(6h_2+p(p-3))+4c^2(f_1^2-f_2-h_2+2p-p^2))=0.\ee 
This has three solutions: one is $p=0$, and the others are
 \be \label{eq:solnppm} p_{\pm}=\frac{(c^2-6)\pm
\sqrt{(c^2-6)^2-16c^2 \Delta}}{2(c^2-2)},
 \ee
 where we have defined \be\Delta=f_1^2-f_2-(1-6/c^2)h_2.\ee The case $p=0$ deserves special mention since in this case (\ref{eq:ABp_1}) and (\ref{eq:ABp_2}) imply that
 $B'=0$. A more careful treatment shows that for this mode $\zeta_k$ and $\delta \sigma_k$ do not behave as the same power for small $\eta$ and $\zeta_k =A'(1+O(k^2\eta^2))$,
 $\delta \sigma_k = A'(0+O(k^2\eta^2))$. 
This mode may be interpreted as the usual constant
$\zeta_k$ mode that appears in single field models. Since $\delta \sigma_k$ vanishes faster that $\zeta_k$, this mode is entirely adiabatic at long wavelengths.

Allowing for all four powers and matching the fields and derivatives at horizon crossing we obtain the following approximate
solution for the positive frequency modes (similar expressions apply for both modes {\bf I} and {\bf II})
\ba \zeta^+_k &=& a_1\, k^{((c^2-6)/(c^2-2)-1/2+\hat{p})} (-\eta)^{(c^2-6)/(c^2-2)} + a_2 \,k^{(-1/2+\hat{p})} \nonumber \\ &&+ a_3 \, k^{(p_{+}-1/2+\hat{p})} (-\eta)^{p_+} + a_4 \, k^{(p_{-}-1/2+\hat{p})} (-\eta)^{p_-},\\
\delta \sigma^+_k&=& -\frac{a_1 c^2(c^2-6)f_1}{2(-6h_2+c^2(f_2+h_2))}\, k^{((c^2-6)/(c^2-2)-1/2+\hat{p})}
(-\eta)^{(c^2-6)/(c^2-2)}
\\ \nonumber &&-\frac{(c^2-2)}{2f_1}\((0+O(k^2\eta^2))a_2 \,k^{(-1/2+\hat{p})}+a_3 p_+ \, k^{(p_{+}-1/2+\hat{p})} \eta^{p_+} + a_4 p_- \,
k^{(p_{-}-1/2+\hat{p})} \eta^{p_-} \),\ea where $a_1 \approx a_2 \approx a_3 \approx a_4 \approx (-\eta_0)^{\hat{p}}$. As a
result the exact adiabatic and isocurvature spectral indices of the modes that dominate at long-wavelengths are given by the
minimum of \be\label{eq:nSposs} n_s={\rm Min} \(\frac{3 c^2-2}{c^2-2} \, , \, \frac{5c^2-14}{c^2-2} \, , \,
4-\frac{1}{|c^2-2|}\sqrt{(c^2-6)^2-16c^2\Delta} \). \ee
The first entry corresponds to the constant $\zeta_k$ mode, the second to the solution described in (\ref{eq:wsoln1}) and (\ref{eq:wsoln2}), and the final term to the $p_{\pm}$ solution given in (\ref{eq:solnppmA}) and (\ref{eq:solnppm}).

It is now apparent that the third term in brackets can be made arbitrarily close to $n_s=1$ by adjusting the field space metric
and potential couplings to give the appropriate value of $\Delta$. In particular exact scale invariance is achieved for \be
\Delta=-\frac{1}{2}(c^2-3)=-\frac{3}{2}w.\ee  However, for this term to be dominant at long wavelengths we require the other two
modes to give $n_s>1$. For the first term this implies $c^2>2$ or $w>-1/3$. Consequently we immediately conclude that this
last term can only be dominant and scale-invariant for contracting solutions. For the second term to be subdominant we require
$c^2>3$ and so $w>0$.

In summary, there are three different ways of achieving scale-invariance, depending on which entry in (\ref{eq:nSposs}) is dominant:
\begin{itemize}
\item The first term in (\ref{eq:nSposs}), at $c^2=0$, $w=-1$. This is an expanding, nearly de Sitter, inflationary universe.  In this case the final term in (\ref{eq:nSposs}) is also scale-invariant any $\Delta$.
\item The second entry in (\ref{eq:nSposs}), at $c^2=3$, $w=0$ provided also that 
$\Delta>0$. This is a contracting solution previously discussed elsewhere.
\item The final entry, for a contracting background with $w>0$ and $\Delta<0$.
\end{itemize}
 In the next section we shall see that the
condition that the contracting scaling solutions are attractors is $\Delta
> 0$ and $w>1$. Therefore none of the contracting backgrounds that give scale-invariant adiabatic perturbation spectra are
stable.

The adiabatic power spectrum for the case where the third term dominates can be expressed in a form similar to that for slow roll
inflationary models \be P_{\zeta}(k) \approx \frac{1}{c^2}\(\frac{k_{phys}}{H_{end}}\)^{2p_-}\frac{H_*^2}{M_{pl}^2}, \ee up to a
factor typically of $O(1)$\footnote{It is feasible that this factor which depends on $c$, $f_1$ and $\Delta$ could be tuned to be
arbitrarily large or small.}, where $H_{*}$ is the Hubble constant at horizon crossing, $H_{end}$ the Hubble constant at the end
of the generation process (\emph{i.e.} reheating) and $k_{phys}$ is the physical wavenumber $k/a(\eta_f)$ at the end of the
generation process. The ratio of isocurvature to adiabatic power is given by \be \frac{P_{\delta
\sigma}}{P_{\zeta}}=\frac{\((c^2-6)- \sqrt{(c^2-6)^2-16c^2 \Delta}\)^2}{16f_1^2} , \ee and so depends on the three parameters
$(c,\Delta,f_1)$. By taking $f_1$ to be suitably large whilst keeping $(c,\Delta)$ fixed we can put an arbitrary fraction of the
power into the adiabatic spectrum without affecting the adiabatic spectral index. Therefore it is not necessary to create
large-amplitude isocurvature fluctuations in order to generate scale-invariant adiabatic fluctuations. In any event the amount of
isocurvature modes that will be observed in the cosmic microwave background depends on the details of the reheating process, or
its analogue in a contracting model.

\section{Stability}\label{s:Stability}

We next determine whether the background scaling solutions are stable to small homogeneous perturbations. Solutions that are
stable have the attractive property that they do not require their initial conditions to be fine-tuned.  The stability properties
are also intimately connected with the spectral indices of long-wavelength perturbations, since the stability analysis is
cosmological perturbation theory in the $k/a \to 0$ limit.  The stability of selected two-field models, with some specific
choices of field space metric and potential, has been studied before \cite{DiMarco:2002eb,Koyama:2007mg}.

In our analysis the conditions required for scale-invariant long-wavelength perturbations will reappear as conditions on the
stability of background solutions.  These conditions can be conveniently expressed in terms of the parameters $c$ and $\Delta$
that were introduced in the preceding section. In the expanding case, the scaling solutions that produce nearly scale-invariant
adiabatic perturbations are always stable.  In the contracting case, the relevant solutions are always unstable.

To begin, we write the equations of motion for the homogeneous components of the fields as an autonomous dynamical system, and then investigate the stability of its fixed points.  The original equations of motion are
given in (\ref{eq:BackgroundEOM1}-\ref{eq:BackgroundEOM3}).  For the time being we will assume $V_0 < 0$ and $c > \sqrt{6}$. We
then define \be (x,y,z,w) = \(\frac{f^{1/2} a \phi'}{\sqrt{6}  a'} \, , \, -\frac{(-V_0 h)^{1/2} a^2e^{-c \phi / 2}}{\sqrt{3} a'}
\, , \, \sigma \, , \, \frac{a \sigma'}{\sqrt{6} a'}\). \ee These dimensionless variables, with the exception of $z$, express the
fractional contribution of each term to the total energy density. The Friedmann equation (\ref{eq:BackgroundEOM1}) now appears as
the constraint equation \be\label{eq:FriedmannConstraint} x^2 + w^2 - y^2 = 1. \ee Using the scalar field equations of motion we
find
\begin{subequations}
\begin{align}
\frac{dx}{d\ln a} & = 3xy^2 - \sqrt{\frac{3}{2}} \frac{1}{f} \frac{df}{d\sigma} x w -  \sqrt{\frac{3}{2}} \frac{c }{f^{1/2}} y^2, \label{eq:AutoSystemXprime}\\
\frac{dy}{d\ln a} & = 3 y(1+y^2) -\sqrt{\frac{3}{2}} \frac{c}{f^{1/2}} xy + \sqrt{\frac{3}{2}} \frac{1}{h} \frac{dh}{d\sigma}wy,  \label{eq:AutoSystemYprime}\\
\frac{dz}{d\ln a} &= \sqrt{6} w, \label{eq:AutoSystemZprime}\\
\frac{dw}{d\ln a} &= 3wy^2 + \sqrt{\frac{3}{2}} \frac{1}{f} \frac{df}{d\sigma} x^2 + \sqrt{\frac{3}{2}} \frac{1}{h} \frac{dh}{d\sigma} y^2.
\label{eq:AutoSystemWprime}
\end{align}
\end{subequations}
As required (\ref{eq:AutoSystemXprime}-\ref{eq:AutoSystemWprime}) propagate the constraint (\ref{eq:FriedmannConstraint}). There is a fixed point at\footnote{Note if $f_{,\sigma}=0$ there are additional fixed points at $(\pm 1,0,z_0,0)$ with $z_0$
constant, however these are physically uninteresting.} \be (x_0,y_0,z_0,w_0) = \(\frac{c}{\sqrt{6f}} \, , \,\sqrt{\frac{c^2}{6f}
- 1} = \sqrt{- \frac{ h\,df/d\sigma }{ f\,dh/d\sigma } \frac{c^2}{6f}}\, ,\, 0 \, ,\, 0\), \ee
which corresponds to the scaling solution studied in previous sections.
The values of $x_0$ and $y_0$
express the requirement that a scaling solution must maintain a constant ratio between the potential and kinetic energy of the
scalar field, which is itself determined by $c$.

To determine whether the fixed point is stable, we expand around it, taking $x = x_0 + \delta x$, \emph{etc}.  The constraint
(\ref{eq:FriedmannConstraint}) implies that
\be
x_0\frac{dx}{d\ln a} =  y_0 \frac{dy}{d\ln a},
\ee
 and so we solve the constraint and eliminate $\delta
y$ as an independent variable by setting \be \delta y = \frac{c }{\sqrt{c^2-6}} \delta x. \ee With this we have three equations
for the perturbations, consistent with the constraint, which are
\begin{subequations}
\begin{align}
\frac{dx}{d\ln a} &= \frac{c^2-6}{2} \delta x - \frac{c f_1}{2} \delta w + \frac{\sqrt{6} c f_1 (c^2 -6)}{24} \delta z ,\\
\frac{dw}{d\ln a} &= -\frac{6 c f_1}{c^2 - 6}\delta x + \frac{c^2 - 6}{2} \delta w - r\delta z ,\\
\frac{dz}{d\ln a} &= \sqrt{6} \delta w,
\end{align}
\end{subequations}
where we have defined \be r = \frac{c^2}{\sqrt{6}(c^2-6)} \left[ c^2 (f_1^2 - f_2 - h_2) - \frac{36}{c^2} h_2 - 3(f_1^2 - 2[f_2 +
2h_2]) \right]. \ee The equations of motion for the perturbations are in the form \be \frac{d\delta x_j}{d \ln a} =  {M_j}^k
\delta x_k, \ee and so the stability of the fixed point is entirely determined by the eigenvalues of ${M_j}^k$.  These are
\begin{subequations}
\begin{align}
\lambda_0 &= \frac{c^2-6}{2} ,\\
\lambda_\pm &= \frac{c^2-6}{4} \left( 1 \pm \sqrt{1 - \frac{16c^2}{(c^2-6)^2} \Delta}\right),
\end{align}
\end{subequations}
where as before \be \Delta = f_1^2 -f_2 - \left[ 1 - \frac{6}{c^2} \right] h_2. \ee The eigenvectors corresponding to
$\lambda_\pm$ are \be -2cf_1 \delta x + (c^2-6)\left[ 1\pm \sqrt{1-\frac{16c^2}{(c^2-6)^2}\Delta} \right] \delta w  +
4\sqrt{6}\delta z, \ee and the eigenvector corresponding to $\lambda_0$ is \be \left[-3(-6h_2 + c^2[f_2 + h_2]) -
c^2(c^2-3)\Delta \right]\delta x +3 (c^2-6) \delta w + 6\sqrt{6} c f_1 \delta z. \ee For a contracting solution, in which $a \to
0$, all of the eigenvalues should be \emph{positive} to ensure stability.  The expression for $\lambda_0$ thus requires \be c >
\sqrt{6} \qquad \text{stable, contracting case,} \ee in accord with previous results with a single scalar field
\cite{Gratton:2003pe}.  With this, $\lambda_+$ is always positive, but $\lambda_-$ is positive only when \be \Delta > 0 \qquad
\text{stable, contracting case}. \ee To study the expanding case, one must start from the original equations of motion but assume
$V_0 > 0$, as required by the scaling solutions with $c<\sqrt{6}$. The corresponding autonomous system, and constraint, can be
derived more simply from the analytic continuation $y \to iy$ in (\ref{eq:FriedmannConstraint}) and
(\ref{eq:AutoSystemXprime}-\ref{eq:AutoSystemWprime}).  The eigenvalues of the matrix ${M_j}^k$ are precisely the same, but since
$a$ is increasing we require the eigenvalues to be negative in order to obtain a stable solution.  This results in the conditions
\be c <\sqrt{6} \qquad \text{stable, expanding case}, \ee again in accordance with results for a single scalar field.  In
addition the expression for $\lambda_-$ requires that \be \Delta
> 0 \qquad \text{stable, expanding case}, \ee to ensure stability of the expanding solution.

Therefore, by comparing these results to those of the previous section, we see that the expanding universes that produce nearly
scale-invariant adiabatic spectra are always stable background solutions.  Furthermore, the contracting universes are always
unstable.  Within this framework it is also possible to study contracting universes with $f_1=0$ that produce scale-invariant
isocurvature perturbations that are later converted into adiabatic modes.  One must carefully redo the analysis since, as
described in the last section, the decoupled $f_1 = 0$ case is not necessarily the limit $f_1 \to 0$ of the coupled case.  In the
end, one finds that the stability properties are controlled by the analogue of $\Delta$ with $f_1$ set to zero. Since the
isocurvature spectral index is smooth in the $f_1 \to 0$ limit, by comparing the results of this section to those of the previous
one, we can conclude that contracting background solutions that produce a scale-invariant spectrum in isocurvature modes alone
are unstable.  Whether this is a problem depends on the specific model one considers.

\section{Conclusions}\label{s:Conclusions}

In the contracting case, we have found that a nearly scale-invariant spectrum of perturbations in the variable $\zeta$ can be
generated continuously as modes exit the horizon for a one-parameter family of $(c,\Delta)$.  The problem of generating such a
spectrum in models where primordial perturbations are laid down during a contracting phase has posed a significant challenge, as
well as sparking lively debate. Other proposed mechanisms to generate the proper spectrum involve a two-stage process, by which
isocurvature perturbations with the proper spectral index are generated on horizon exit, and then later converted into adiabatic
perturbations. The mechanism we describe here is more akin to the classic inflationary mechanism in which the desired density
perturbation is created in a single step as modes exit the horizon.  Furthermore, this can be achieved entirely within a
four-dimensional effective theory, without recourse to higher-dimensional effects or higher-derivative interactions.  There
remains the challenge of a smooth big crunch/big bang transition, but assuming that this is possible, it seems that within the
context of a simple model it is certainly possible to generate the required spectrum of primordial density perturbations in
cosmological models with a collapsing phase.

In the expanding case, we find the expected conclusion that a nearly scale-invariant perturbation spectrum can only be obtained
for $w \approx -1$, i.e. close to slow roll, regardless of how we tune the other couplings $(f_1, \Delta)$. This is because there
always exists a constant mode in $\zeta$ which necessarily dominates the long-wavelength power spectrum and whose spectral index
is solely determined by the equation of state $w$.

In this work we have obtained these results by studying idealized two scalar systems that admit scaling solutions.  The scaling
symmetry is powerful enough to enable the background evolution and the spectral indices of perturbations to be computed exactly.
The scaling symmetry is also general enough to provide reasonable representatives of a large class of models in which the
equation of state parameter $w$ varies slowly during the time that primordial perturbations are laid down.  Within the context of
these scaling solutions it has been possible to completely characterize the conditions under which a scale-invariant spectrum of
density perturbation is created, after the spirit of previous studies in the one-field case
\cite{Lyth:1991bc,Gratton:2003pe,Boyle:2004gv}.

There are several directions in which the work presented here should be extended.  With respect to model building, the
possibility of achieving a scale-invariant adiabatic perturbation spectrum with a simple two-derivative action
 opens up new avenues for realizing cosmological models with a contracting phase.
This is especially promising with respect to string theory, where multiple scalar fields and curved field spaces are common.
Thanks to the instability in the contracting solutions, complete cosmological models that incorporate this mechanism
 must explain how the scalar fields begin with the proper initial conditions in field space. 


\section*{Acknowledgements}
We would like to thank Latham Boyle, Claudia de Rham, Justin Khoury, Jean-Luc Lehners, Katie Mack, Joao Magueijo, Paul
Steinhardt, Bartjan van Tent, Neil Turok and especially Krzysztof Turzynski for useful discussions. DHW thanks the Perimeter
Institute for its hospitality during the completion of this work. Research at Perimeter Institute for Theoretical Physics is
supported in part by the Government of Canada through NSERC and by the Province of Ontario through MRI.

\end{document}